\documentclass[11pt]{article}

\usepackage{amsmath,amssymb,latexsym,color}
\usepackage{multicol}
\usepackage{graphicx}
\usepackage{amsfonts}

\newcommand{\rem}[1]{}
\newcommand{\remfigure}[1]{}

\DeclareMathAlphabet{\mathbi}{OML}{cmm}{b}{it} 

\makeatletter
\def\extrarowheight#1{\noalign{\@tempdima\ht\@arstrutbox
\advance\@tempdima#1\ht\@arstrutbox\@tempdima}}
\makeatother

\newcommand{\bel}{\begin{equation}\label}
\newcommand{\ee}{\end{equation}}
\newcommand{\ben}{\begin{enumerate}}
\newcommand{\een}{\end{enumerate}}
\newcommand{\bx}{\mbox{\boldmath$x$}}
\newcommand{\by}{\mbox{\boldmath$y$}}

\newcommand{\bG}{Gr}
\newcommand{\Rey}{Re}

\newcommand{\bdf}{\mathbi{f}}

\newcommand{\bu}{\mathbi{u}}

\newcommand{\bv}{\mathbi{v}}

\newcommand{\bphi}{\mbox{\boldmath$\Phi$}}

\newcommand{\beq}{\begin{eqnarray}\label} 
\newcommand{\eeq}{\end{eqnarray}} 
\newcommand{\bc}{\begin{center}} 
\newcommand{\ec}{\end{center}} 
\newcommand\shalf{\ensuremath{{\scriptstyle\frac{1}{2}}}}
\newcommand{\I}{\int_{\Omega}}
\newcommand{\lans}{LANS-$\alpha$~}
\newcommand{\etal}{\textit{et al.~}}
\newtheorem{theorem}{Theorem}
\newtheorem{lemma}{Lemma}

\makeatletter
\@addtoreset{equation}{section}

\makeatother

\pagecolor{white} 
\color{black} 

\begin{document}
\bc
\textbf{\Large Length-scale estimates for the \lans equations\\ 
in terms of the Reynolds number}
\ec
\bc
{\large J. D. Gibbon and D. D. Holm}
\par\vspace{2mm}
Department of Mathematics, Imperial College London, London SW7 2AZ, UK\,.
\ec
\begin{abstract}
Foias, Holm \& Titi \cite{FHT2} have settled the problem of existence and uniqueness for 
the $3D$ \lans equations on periodic box $[0,L]^{3}$. There still remains the problem, 
first introduced by Doering and Foias \cite{DF} for the Navier-Stokes equations, of 
obtaining estimates in terms of the Reynolds number $\Rey$, whose character depends on 
the fluid response, as opposed to the Grashof number, whose character depends on the 
forcing. $\Rey$ is defined as $\Rey = U\ell/\nu$ where $U$ is a bounded spatio-temporally 
averaged Navier-Stokes velocity field and $\ell$ the characteristic scale of the forcing.
It is found that the inverse Kolmogorov length is estimated by $\ell\lambda_{k}^{-1} 
\leq c\,(\ell/\alpha)^{1/4}\Rey^{5/8}$. Moreover, the estimate of Foias, Holm \& Titi for 
the fractal dimension of the global attractor, in terms of $\Rey$, comes out to be
$$
d_{F}(\mathcal{A}) \leq c\,
\frac{V_{\alpha}V_{\ell}^{1/2}}{(L^{2}\lambda_{1})^{9/8}}\,\Rey^{9/4}\,
$$
where $V_{\alpha} = \left(L/(\ell\alpha)^{1/2}\right)^{3}$ and $V_{\ell} = 
\left(L/\ell\right)^{3}$.
It is also shown that there exists a series of time-averaged inverse squared 
length scales whose members, $\left<\kappa_{n,0}^2\right>$, 
are estimated as $(n\geq 1)$
$$
\ell^{2}\left<\kappa_{n,0}^2\right> \leq c_{n,\alpha}V_{\alpha}^{\frac{n-1}{n}}
\Rey^{\frac{11}{4} - \frac{7}{4n}}(\ln\Rey)^{\frac{1}{n}} + c_{1}\Rey(\ln\Rey)\,.
$$
The upper bound on the first member of the hierarchy $\left<\kappa_{1,0}^2\right>$ 
coincides with the inverse squared Taylor micro-scale to within log-corrections.
\end{abstract}

\section{\large Introduction}\label{mainintro}

\subsection{Background to the LANS-$\alpha$ model}\label{background}

Turbulence is a state of continual unrest, which arises as a fluid's response to forcing by 
stirring, for example, or by flowing along a wall. The scaling properties of turbulence 
are characterized by two dimensionless numbers. These are the Grashof number $\bG$, which 
measures forcing, and the Reynolds number $\Rey$ which characterizes the fluid's response 
to the forcing. The turbulent response to forcing produces strong fluctuations in the fluid 
motion whose statistics obey power law spectra extending over a large range of length scales 
and time scales \cite{Ko1941,Frisch}. This fluctuating multi-scale response is the hallmark 
of turbulence. Turbulence researchers often characterize the development of the multi-scale 
response as a ``cascade'' of kinetic energy rushing downward from the larger fluid motions 
due to forcing to the smaller and smaller circulations of  eddies, sheets, and tubes of
vorticity. In  stretching themselves into finer and finer shapes, these vortical structures 
comprise the ``sinews'' of turbulence.

Characteristic features of turbulence -- its distribution of eddy sizes, shapes, speeds, 
vorticity, circulation and viscous dissipation -- may all be captured by using the exact 
Navier-Stokes equations. These correctly predict how the cascade of turbulent kinetic energy 
and vorticity accelerates and how the sinews of turbulence stretch themselves into finer and 
finer scales, until their motions reach only a few molecular mean free paths, where they may 
finally be dissipated by viscosity into heat\,: for more details, see 
\cite{Frisch,CF,Temam,Fo1997,FoTe1987,ViFu1980}. However, the fidelity of the Navier-Stokes 
equations in capturing the cascade of turbulence is also the cause of serious problems 
in direct numerical simulations of turbulence.

Based on Landau's heuristic idea that the number of active degrees of freedom required to 
simulate the turbulent cascade in high-Reynolds-number flows increases as $\Rey^{9/4}$, it 
is clear that this geometric rate of increase quickly outstrips the numerical resolution 
capabilities of even the largest computer. To make turbulence computable, scientists have 
developed various approximate models that halt the cascade into smaller, faster eddies. In 
most models, this effect is accomplished by causing the eddies below a certain size to 
dissipate computationally into heat. This dissipative imperative causes errors, however, 
because it damps out the variability (known as intermittency) in the larger-scale flow, 
which is caused by the myriad of small scales of motion interacting nonlinearly together 
in the fields of the larger motions. Thus, computational turbulence closure models based 
on enhancing viscous dissipation over its physical Navier-Stokes value run the risk of 
producing unrealistically low variability.  

Perhaps surprisingly, one of the first clues in understanding how to develop turbulence closure 
models without enhancing viscous dissipation came from the great mathematical analyst Leray 
\cite{Leray} who showed how to regularize the Navier-Stokes equations by modifying their 
nonlinearity to the well-known form
\bel{NS1}
\bv_{t} + \bu\cdot\nabla\bv + \nabla p = \nu\Delta\bv + \bdf(\bx)
\,,\quad
\rm{div}\,\bv = 0\,,
\ee
with $\bv = 0$ on the boundary. Here $\nu$ is the (constant) kinematic viscosity coefficient, 
$\bdf(\bx)$ is the prescribed  external force and $\bu = G_\delta*\bv$ is a filtered version of 
the regularized velocity $\bv$. The filtering operation is defined by $G_{\delta}*\bv = 
\int G_{\delta}(\bx,\,\by)\bv(\by)\,d\,^3y$ for a symmetric kernel $G_\delta(\bx,\,\by)$ of 
characteristic width $\delta$. The Navier-Stokes equations for $\bv$ are recovered in 
the limit as $\delta\to0$, so that $\bu \to\bv$. For a review of the Leray regularization of 
the Navier-Stokes equations, see Gallavotti \cite{Ga1993}.

One of the points made in \cite{Ga1993} is that the Leray regularization of the Navier-Stokes 
equations no longer satisfies the Kelvin circulation theorem, since for these equations
\bel{NS2}
\frac{d}{dt}\oint_{\Gamma(\bu)}\bv\cdot d\bx = \oint_{\Gamma(\bu)}\Big(\bv_{t} + \bu\cdot\nabla\bv +
\nabla\bu^{T}\cdot\bv\Big)\cdot d\mathbf{x} \ne \oint_{\Gamma(\bu)}\Big(\nu\Delta\bv + \bdf\Big)
\cdot d\bx\,.
\ee
Remarkably, combining the process of Lagrangian averaging with Taylor's hypothesis (that fluctuations
have such low power that they may be regarded as being carried along by the mean flow) leads to a 
regularized set of equations which answers the challenge of \cite{Ga1993} to produce a regularization
of the Navier-Stokes equations which do possess a Kelvin circulation theorem. These regularized equations
comprise the \lans model\footnote{LANS is an acronym standing for ``Lagrangian-averaged Navier-Stokes''
while alpha $(\alpha)$ is the coherence length of the Lagrangian statistics; this term will be used
hereafter.}
\bel{uv2a}
\bv_{t} + \bu\cdot\nabla\bv + \nabla\bu^T\cdot\bv + 
\nabla P = \nu\Delta\bv + \bdf(\bx)\,,\quad \rm{div}\,\bu = 0 \,,
\ee
where $P = p - \shalf\nabla\big(|\bu|^2 + \alpha^{2}|\nabla\bu|^2\big)$.
An equivalent alternative formulation is to rewrite (\ref{uv2a}) as
\bel{uv2b}
\bv_{t}- \bu\times\mbox{curl}\,\bv = \nu\Delta\bv - \nabla \tilde{p} +\bdf(\bx)
\ee
where $\tilde{p} = P + \bu\cdot\bv$. Usually, $\alpha$ is taken as a constant with dimension of length and the filtering relation $\bu = G_{\alpha}*\bv$ for the advection velocity in the LANS$-\alpha$ model  is specified as
\bel{uv1}
\bv \equiv \bu - \alpha^2\Delta\bu\,. 
\ee
Some remarks are in order here:
\begin{enumerate}
\item The filtering kernel $G_\alpha$ for the LANS$-\alpha$ model thus turns out to 
be the Green's function for the Helmholtz operator $(1-\alpha^2\Delta)$. 

\item As expected, the LANS$-\alpha$ motion equation satisfies the Kelvin circulation
theorem:
\bel{KThm}
\frac{d}{dt}\oint_{\Gamma(\bu)} \bv \cdot d\bx =
\oint_{\Gamma(\bu)}\Big(\bv_{t} + \bu\cdot\nabla\bv +
\nabla\bu^T\cdot\bv\Big)\cdot d\mathbf{x} = \oint_{\Gamma(\bu)}
\Big(\nu\Delta\bv + \bdf\Big)\cdot d\bx\,.
\ee
The circulation theorem tells us that the rate of change of circulation of
momentum per unit mass $\bv$ around a closed material loop $\Gamma(\bu)$ moving 
with velocity $\bu = G_\alpha*\bv$ is due to the integral around that loop of the 
tangential component of the sum over forces (viscous and external) acting on the fluid. 

\item This statement of the circulation theorem is also a mnemonic for deriving other 
regularized turbulence models of LANS$-\alpha$ type by specifying a different 
filtering kernel $G_\alpha$ \cite{GeHo2003}.
\end{enumerate}
We have seen that the LANS$-\alpha$ model can be immediately derived from its circulation 
theorem. However, the approach used historically in Chen \etal \cite{Chen-etal1998} 
and Holm and Titi \cite{HT05} for 
deriving the closed Eulerian form (\ref{uv2a}) of the LANS$-\alpha$ motion equation was 
based on the combination of two other earlier results. First, the Lagrangian-averaged 
variational principle of Gjaja and Holm \cite{GjHo1996} was applied for deriving the 
inviscid averaged nonlinear fluid equations, obtained by averaging Hamilton's principle 
for fluids over the rapid phase of their small turbulent circulations at fixed Lagrangian 
coordinate\,: this step had its own precedent in earlier work on Lagrangian-averaged fluid 
equations by Andrews and McIntyre \cite{AnMc1978}. Second, the Euler-Poincar\'e theory for 
continuum mechanics of Holm, Marsden and Ratiu \cite{HoMaRa1998} was used for handling the 
Eulerian form of the resulting Lagrangian-averaged fluid variational principle. This second 
step determined the relationships among the momentum per unit mass $\bv$, the Lagrangian-averaged 
velocity of the fluid  $\bu$ and the Lagrangian fluctuation statistics. Next, Taylor's hypothesis 
of frozen-in turbulence circulations was invoked for closing the Eulerian system of Lagrangian-averaged
fluid equations, by obtaining the explicit relation $\bv \equiv \bu - \alpha^2\Delta\bu$.  Finally,
Navier-Stokes Eulerian viscous dissipation was added, so that viscosity would cause diffusion of 
the newly-defined Lagrangian-average momentum and monotonic decrease of its total 
Lagrangian-averaged energy. 

At this point, one may regard the LANS$-\alpha$ model as an alternative regularization of the 
Navier-Stokes equations and re-examine its properties from the viewpoint of Leray's analysis:  
in fact the problem of existence and uniqueness for the \lans equations has already been settled 
by Foias, Holm \& Titi \cite{FHT2}. In addition, the same ideas which restore Kelvin's circulation 
theorem to Leray's regularization of the Navier-Stokes equations also turn out to provide a basis 
for deriving candidate equations for Large Eddy Simulations (LES) of turbulence. Conversely, other 
proposed LES models of turbulence may lead to likely candidates for application of Leray's analysis.
In this way, the classical Leray analysis of the Navier-Stokes equations finds itself a new role in
the study of the analytical properties of turbulence models. Indeed, the Leray model itself was 
recently found to be a viable candidate for LES computational modeling of turbulence \cite{GeHo2003,CHOT}.

\subsection{\large Estimates in terms of Reynolds number}\label{intro}

Motivated by the Navier-Stokes equations, in an early and progressive paper Ruelle \cite{Ruelle82}
discussed how ideas in dynamical systems might be extended to the infinite dimensional case by 
counting the number of positive characteristic exponents. Certain general finiteness assumptions 
were made about the nature of the Navier-Stokes equations without formally using the concept of a 
global attractor. These ideas 
were taken to another level by Constantin and Foias \cite{CF85} who used the idea of a global 
attractor $\mathcal{A}$ for a partial differential equation to determine the number of growing 
Lyapunov exponents and thence, through a rigorous generalization of the Kaplan-Yorke formula, 
they were able to get formal upper bounds on $d_{L}(\mathcal{A})$, the Lyapunov dimension 
of $\mathcal{A}$, which itself bounds above $d_{F}(\mathcal{A})$ and $d_{H}(\mathcal{A})$. 
Importantly these bounds depend only upon time-averages and thus only on the long-time dynamics 
on $\mathcal{A}$; see also \cite{CFT1}. When applied to the two-dimensional Navier-Stokes 
equations, for which a global attractor exists, these methods gave good, sharp estimates to 
within logarithmic corrections \cite{CFT2} when used in conjunction with an $L^{\infty}$ 
estimate of Constantin \cite{Const1}. The development of the key features of these ideas 
can be found in \cite{CF,IDDS}, including the role played by the inequalities of Lieb and 
Thirring \cite{LT}. Apart from the two-dimensional 
Navier-Stokes equations, several important partial differential equations possess a global 
attractor, such as the the two-dimensional complex Ginzburg-Landau equation and the one-dimensional
Kuramoto-Sivashinky equation. Unfortunately, the three-dimensional Navier-Stokes equations are 
not among them; the key element is the lack of a proof of existence and uniqueness without which 
the existence of a global attractor remains open. 

In settling the question of existence and uniqueness for the \lans equations, Foias, 
Holm \& Titi \cite{FHT2} were able to use the machinery developed for global attractors 
to find estimates for the dimension of its global attractor $\mathcal{A}$ (and other important 
quantities) in terms of the Grashof number $\bG$, which is a dimensionless control parameter 
dependent only on the ratio of the forcing to the viscosity $\nu$. Therefore, as a regularization 
of the three-dimensional Navier-Stokes equations with many similar features to its parent, the 
\lans equations equations possess this extra key property. \textit{What remains to be proved, 
however, is whether the estimates in \cite{FHT2} can be evaluated in terms of the 
Reynolds number, whose character depends on the fluid response to the forcing, and which is 
intrinsically a property of Navier-Stokes solutions.} The advantage of this further step lies 
in the fact that the engineering and physics communities express their ideas about 
turbulence in terms of the Reynolds number. It also allows us to make direct comparisons 
between estimates for the two equations where they exist (see Table 1).

For simplicity the \lans equations (\ref{uv2a}), or their alternative form in (\ref{uv2b}), 
will be considered on a periodic domain $[0,\,L]^{3}_{per}$ with forcing $\bdf(\bx)$ taken 
to be $L^2$-bounded of narrow-band type\footnote{The restriction to narrow-band forcing can 
be relaxed at the cost of more parameters in the problem.} with a single length-scale $\ell$ 
(see \cite{DF,DG02,GD05})
\bel{f1}
\|\nabla^n\bdf\|_{2} \approx \ell^{-n}\|\bdf\|_{2}\,. 
\ee
With $f_{rms} = L^{-d/2}\|\bdf\|_{2}$, where and $\|\bdf\|^{2}_{2} = \int_{\Omega}|\bdf|^2\,dV$, 
the standard definition of the Grashof number in $d$-dimensions is
\bel{f2}
\bG = \frac{\ell^{3}f_{rms}}{\nu^2}\,.
\ee
Analytical estimates in Navier-Stokes theory have traditionally been expressed in terms 
of $\bG$ \cite{CF,Temam,Lady,Serrin,Foiasstat1,Foiasstat3,DGbook} but these are difficult 
to compare with the results of Kolmogorov scaling theories  which are expressed in terms 
of Reynolds number \cite{Frisch}. A good definition of this is
\bel{Redef}
\Rey = \frac{U\ell}{\nu}
\hspace{3cm}
U^{2} = L^{-d}\left<\|\bu\|^{2}_{2}\right>
\ee 
where $\left<\cdot\right>$ is the long-time-average
\bel{tadef}
\left<g(\cdot)\right> = \overline{\lim}_{t\to\infty}\frac{1}{t}
\int_{0}^{t}g(\tau)\,d\tau\,.
\ee
Doering and Foias \cite{DF} addressed this problem recently and have shown that in the 
limit $\bG\to\infty$, solutions of the $d$-dimensional Navier-Stokes equations 
\bel{NSequns}
\bu_{t} + \bu\cdot\nabla\bu + \nabla p = \nu\Delta\bu + \bdf(\bx)
\,,\quad
\rm{div}\,\bv = 0\,.
\ee
must satisfy\footnote{This result is not advertised in \cite{DF} but it follows immediately 
from their equation (48). \cite{DF} also contains another result that the energy dissipation 
rate $\epsilon$ has a lower bound proportional to $\bG$: see Appendix (\ref{App1.2}).}
\bel{DF1}
\bG \leq c\,(\Rey^{2} + \Rey) \,.
\ee
\vspace{-3mm}
\begin{center}
\begin{tabular}{||c|c|c|c||}
\hline\hline
\extrarowheight{12mm}
Time-average & NS & \lans & Eqn number \\\hline\hline
$\ell\lambda_{k}^{-1}$ & $\Rey^{3/4}$ & $\Rey^{3/4}$ & (\ref{energy3})\\ \hline
$\ell\lambda_{k}^{-1}$  (improved) & unknown & $(\ell/\alpha)^{1/4} \Rey^{5/8}$ & (\ref{uv5d})\\ \hline
$\alpha^{2}\ell\nu^{-2}\left<H_{2}\right>$ & unknown & $V_{\ell}\,\Rey^{3}$ & (\ref{uv5b1})\\ \hline
$d_{F}(\mathcal{A})$ & unknown & $V_{\alpha}V_{\ell}^{1/2}(L^{2}\lambda_{1})^{-9/8}
\,\Rey^{9/4}$ & (\ref{ad6})\\\hline
$\ell^{2}\left<\kappa_{n,0}^2\right>$ & unknown & $V_{\alpha}\,\Rey^{\Lambda_{n}}(\ln\Rey)^{\frac{1}{n}}$
&(\ref{th2a})\\\hline$\ell^{2}\left<\kappa_{1,0}^2\right>$ & $\Rey(\ln\Rey)$ & $\Rey(\ln\Rey)$ & (\ref{th2b})\\\hline
$\ell^{2}\nu^{-2}\left<\|\bu\|_{\infty}^2\right>$ & unknown &$V_{\alpha}\,\Rey^{11/4}$&(\ref{th2c})\\\hline
$\alpha\ell\nu^{-1}\left<\|\nabla\bu\|_{\infty}\right>$ & unknown & 
$V_{\alpha}^{1/4}V_{\ell}^{1/2}\,\Rey^{35/16}$ & (\ref{th2d})\\\hline
$\ell\left<\kappa_{n,0}\right>$ & $V_{\ell}\,\Rey^{3-5/2n}(\ln\Rey)^{\frac{1}{n}}\,$ 
& $V_{\alpha}^{1/2}\,\Rey^{\Lambda_{n}/2}(\ln\Rey)^{\frac{1}{2n}}$ & (\ref{th2a})\\\hline
\hline
\end{tabular}\label{tab1e1} 
\end{center}
\vspace{0.15cm}
\par\medskip\noindent
Table 1: {\small Comparison of various time-average bounds for the Navier-Stokes and \lans 
equations with constants omitted.  Note that in the cases where there is no known equivalent 
upper bound for the Navier-Stokes equations, the \lans upper bounds blow up as $\alpha\to 0$: 
$\Lambda_{n}$  is defined in (\ref{Vdef}). Lines 5-8 are a summary of the results of Theorem 
\ref{thm2} in \S\ref{highderiv}. The Navier-Stokes estimate for $\ell\left<\kappa_{n,0}\right>$ 
on the last line is there for comparative purposes and is taken from \cite{DG02,GD05}.}
\par\medskip
Using the relation in (\ref{DF1}), Doering and Gibbon \cite{DG02,GD05} have re-expressed 
some $3D$ Navier-Stokes estimates in terms of $\Rey$ (see Table 1). The problem, however, 
is less simple than substituting (\ref{DF1}) into standard results, although this works 
well enough for point-wise estimates \cite{FHT2}. Time averages are more subtle and exploit 
how the average velocity $U$ within $\Rey$ is related to $\left<\|\bu\|_{2}^2\right>$. As 
an illustration, let us consider the Navier-Stokes equations (\ref{NSequns}) whose energy 
dissipation rate is $\epsilon = \nu\left<\|\nabla\bu\|_{2}^{2}\right>L^{-d}$. Standard 
estimates show that its upper bound is proportional to $\bG^{2}$. By (\ref{DF1}), this 
turns into $\Rey^{4}$, which is not sharp.  Now we estimate this a different way \cite{DF}: 
consider Leray's energy inequality
\bel{energy1}
\shalf\frac{d~}{dt} \|\bu\|_{2}^{2} 
\leq -\nu \|\nabla\bu\|_{2}^{2} + \|\bdf\|_{2}\|\bu\|_{2}\,.
\ee
Time-averaging (\ref{energy1}) and using (\ref{Redef}) and (\ref{DF1}) yields
\bel{energy2}
\epsilon \leq \nu^{3}\ell^{-4}\bG\,\Rey
\leq c\,\nu^{3}\ell^{-4}\left(\Rey^{3}+\Rey\right)\,,
\ee
which is a considerable improvement. To leading order the inverse Kolmogorov length 
$\lambda_{k}^{-1} = (\epsilon/\nu^{3})^{1/4}$ is then bounded above by 
\bel{energy3}
\ell\lambda_{k}^{-1} \leq c\,\Rey^{3/4}\,.
\ee
This estimate now conforms with the generally accepted scaling law for the inverse Kolmogorov 
length with the Reynolds number \cite{Ko1941,Frisch}. 

The relation in (\ref{DF1}) is essentially a Navier-Stokes result and thus needs re-proving 
for the \lans equations. In turns out to be true but the proof is a non-trivial extension of 
the method in \cite{DF}; the whole of Appendix (\ref{App1.1}) is devoted to this proof. As 
will be shown in \S\ref{H1ball}, the estimate for $\left<H_{1}\right>$ in (\ref{energy2}) 
can be improved to
\bel{H1improve}
\epsilon \leq c\,\nu^{3}\ell^{-3}\alpha^{-1}\Rey^{5/2}\,,
\hspace{1cm}
\Rightarrow
\hspace{1cm}
\ell\lambda_{k}^{-1} \leq c\,\left(\frac{\ell}{\alpha}\right)^{1/4}\Rey^{5/8}\,.
\ee
Moreover, in \S\ref{attdim} the estimate by Foias, Holm and Titi \cite{FHT2} (see also \cite{FHT1})
for the fractal dimension $d_{F}(\mathcal{A})$ of the global attractor $\mathcal{A}$ is considered 
in the light of these $\Rey$-bounds. Using their estimate in terms of the generalized 
$\alpha$-dependent dissipation rate, we show that 
\bel{nine4}
d_{F}(\mathcal{A}) \leq c\,\frac{V_{\alpha}V_{\ell}^{1/2}}{(L^{2}\lambda_{1})^{9/8}}
\,\Rey^{9/4}\,
\ee
where the two dimensionless volumes $V_{\ell}$ and $V_{\alpha}$ are defined by
\bel{vol}
V_{\ell} = \left(\frac{L}{\ell}\right)^{3}
\hspace{2cm}
V_{\alpha} = \left(\frac{L}{(\ell\alpha)^{1/2}}\right)^{3}\,,
\ee
and $\lambda_{1} > 0$ is smallest eigenvalue of the Stokes operator. Given our definition of 
$\Rey$ in (\ref{Redef}), this $\Rey^{9/4}$ estimate is consistent with scaling theories of 
turbulence but it does not survive in the Navier-Stokes limit because the volume $V_{\alpha}$ 
blows up as $\alpha\to 0$.

The $\Rey^{9/4}$ estimate for $d_{F}$ also gives an idea of how many degrees of freedom, 
in Landau's sense, exist in a turbulent flow -- indeed this is exactly result predicted by 
Landau for the Navier-Stokes equations\footnote{However, given the improved $\Rey^{5/8}$ 
inverse Kolmogorov estimate in (\ref{H1improve}) for the \lans model, it is possible that 
the sharp estimate for $d_{F}$ is proportional to $\Rey^{15/8}$.}. What is not taken into 
account in this picture is the effect of strong dissipation-range intermittency where 
significant energy lies in wave-numbers larger than $\lambda_{k}^{-1}$. In this case 
estimates are needed for length-scales that are associated with higher derivatives.  This 
idea has been investigated in \cite{DG02,GD05} where estimates were found for time-averaged 
Navier-Stokes quantities. Those for the \lans equations should be much better because of 
their enhanced regularity properties \cite{FHT2}. In \S\ref{highderiv} we combine the 
forcing with higher derivatives of the velocity field in the form
\bel{Fdef1a}
F_{n} = H_{n} +\tau^{2} \|\nabla^{n}\bdf\|_{2}^{2}\,,
\ee
where $\tau = \ell^{2}\nu^{-1}(\bG\ln\bG)^{-1/2}$ is a characteristic time\,: see 
Appendix (\ref{App1.2}). We also form the combination
\bel{Jndefa}
J_{n} = F_{n} + \alpha^{2}F_{n+1}
\ee
and use it to define a set of inverse length scales, or time-dependent wave-numbers,
\bel{2nmom1}
\kappa_{n,0}(t) = \left(\frac{J_{n}}{J_{0}}\right)^{\frac{1}{2n}}\,.
\ee
In the $\alpha\to 0$ limit, the $\kappa_{n,0}^{2n}$ behave as the $2n$th-moments of the 
energy spectrum. Theorem \ref{thm2} of \S\ref{highderiv} proves that the time average of 
their squares must obey
\bel{2nmom2}
\ell^{2}\left<\kappa_{n,0}^2\right> \leq c_{n,\alpha}V_{\alpha}^{\frac{n-1}{n}}
\Rey^{\Lambda_{n}}(\ln\Rey)^{\frac{1}{n}} + c_{1}\Rey\ln\Rey\,,
\ee
for the \lans model, where
\bel{Vdef}
\Lambda_{n} = \frac{11}{4} - \frac{7}{4n}\,.
\ee
Note that the $n=1$ estimate in (\ref{2nmom2}) scales with $\Rey$ the same as the Taylor micro-scale.
This best that could be achieved for the full $3D$ Navier-Stokes equations was \cite{DG02,GD05}
\bel{NS1}
\ell\left<\kappa_{n,0}\right>\leq c_{n}V_{\ell}\,\Rey^{3 -\frac{5}{2n}}(\ln\Rey)^{\frac{1}{n}}
+ c_{1}\Rey\ln\Rey\,,
\ee
which appears in the last line of Table 1. With the exponent of unity in the time-average --if 
indeed a solution exists at all --  this 
is not only much worse that (\ref{2nmom2}) but represents only weak solutions, as opposed to the 
strong solutions of Foias, Holm and Titi \cite{FHT2}.  The fact that no upper bound is known to 
exist for $\left<\kappa_{n,0}^2\right>$ for the $3D$ 
Navier-Stokes equations is consistent with the fact that the dimensionless volume $V_{\alpha}$ 
blows up as $\alpha\to 0$. The $\kappa_{n,0}$ could even become singular in this limit. 


\section{\large Properties of the product $\I\bu\cdot\bv\,dV$}\label{H1ball}

Foias, Holm and Titi \cite{FHT2} noted that the product $\bu\cdot\bv$ has two convenient 
properties
\bel{uv3}
\I \bu\cdot\bv\,dV = \I \left\{|\bu|^{2} + \alpha^{2}|\nabla\bu|^{2}\right\}dV
\ee
\vspace{-2mm}
and
\vspace{-2mm}
\beq{uv4}
\frac{d~}{dt}\I \bu\cdot\bv\,dV  &=&  \I(\bu_{t}\cdot\bv + \bu\cdot\bv_{t})\,dV\nonumber\\
&=& \I(\bu_{t}\cdot(1-\alpha^{2}\Delta)\bu + \bu\cdot\bv_{t})\,dV \nonumber\\
&=& \I\left\{\bu\cdot\left[1-\alpha^{2}\Delta)\bu_{t}\right] + \bu\cdot\bv_{t}\right\}\,dV
\nonumber\\
&=& 2\I\bu\cdot\bv_{t}\,dV
\eeq
where two integrations by parts have occurred between the second and third lines. 
Now define
\bel{Hndef}
H_{n} = \I |\nabla^{n}\bu|^2\,dV \equiv  \I |\mbox{curl}^{n}\bu|^2\,dV 
\ee
this being true on a periodic domain because $\mbox{div}\,\bu = 0$. 
Clearly we have the bound
\beq{uv5a}
\shalf\frac{d~}{dt}\left(H_{0} + \alpha^{2}H_{1}\right) 
&=& -\nu \left(H_{1} + \alpha^{2}H_{2}\right) + \I\bu\cdot \bdf\,dV\nonumber\\
&\leq & -\nu \left(H_{1} + \alpha^{2}H_{2}\right) + \|\bu\|_{2}\|\bdf\|_{2}\,.
\eeq
One can then calculate an absorbing ball for $H_{1}$ with ease (see \cite{FHT2}). It 
is also possible to estimate the time averages $\left<H_{1}\right>$ and $\left< H_{2}\right>$ 
which can be found in the same manner as in (\ref{energy1}) to satisfy
\bel{uv5b}
\nu L^{-3}\left<H_{1} + \alpha^{2}H_{2}\right> 
\leq \nu^{3}\ell^{-4}\Rey\,\bG \leq c\,\nu^{3}\ell^{-4}\Rey^{3}\,.
\ee
The upper bound on $\left<H_{2}\right>$, written as
\bel{uv5b1}
\alpha^{2}\ell\nu^{-2}\left<H_{2}\right> \leq c\,V_{\ell}\Rey^{3}\,
\ee
can then be used 
to improve the estimate for $\left<H_{1}\right>$ by using the fact that $\left<H_{1}\right>
\leq \left< H_{0}\right>^{1/2}\left<H_{2}\right>^{1/2}$ and that 
$U^{2} = L^{-3}\left<H_{0}\right>$. We find that 
\bel{uv5c}
\left<H_{1}\right> \leq c\,\nu^{2}L^{3}\ell^{-3}\alpha^{-1}\Rey^{5/2}\,,
\ee
and so
\bel{uv5d}
\ell\lambda_{k}^{-1} \leq c\,\left(\frac{\ell}{\alpha}\right)^{1/4}\Rey^{5/8}\,.
\ee
Hence the energy dissipation rate $\epsilon$ is also bounded above by $\Rey^{5/2}$ but the 
improved estimate blows up when $\alpha\to 0$; no equivalent result is implied for the 
$3D$ Navier-Stokes equations.


\section{\large A $\Rey^{9/4}$ bound for the attractor dimension}\label{attdim}

Foias, Holm and Titi \cite{FHT2} made two independent estimates of the fractal dimension 
$d_{F}(\mathcal{A})$ of the global attractor $\mathcal{A}$. The first was in terms of 
$\bG$ but the second estimate was made in terms of the ``energy dissipation rate'' 
$\overline{\epsilon}$; this phrase has been put in inverted commas because it includes 
the $H_{2}$-norm, whereas conventionally only the $H_{1}$-norm is used. Their definition 
of $\overline{\epsilon}$ is 
\bel{ad1}
\overline{\epsilon} = \lambda_{1}^{3/2}\nu\left<H_{1}+\alpha^{2}H_{2}\right>
\ee
where $\lambda_{1}$ is the smallest eigenvalue of the Stokes' operator which has the 
dimension of an inverse length squared.  Foias, Holm and Titi \cite{FHT2} then proved that 
\bel{ad2}
d_{F}(\mathcal{A}) \leq c\,\frac{\lambda_{1}^{-3/2}}{(\alpha^{2}\lambda_{1})^{3/4}}
\left(\frac{\overline{\epsilon}}{\nu^{3}}\right)^{3/4}\,.
\ee
They then defined a Kolmogorov length as $\ell_{\epsilon}^{-1} = 
\left(\overline{\epsilon}/\nu^{3}\right)^{1/4}$ which turns (\ref{ad2}) into
\bel{ad3}
d_{F}(\mathcal{A}) \leq c\,\frac{1}{(\alpha^{2}\lambda_{1})^{3/4}}
\frac{1}{(\ell_{\epsilon}\lambda_{1}^{1/2})^{3}}\,.
\ee
The problem here lies in interpretation: $\ell_{\epsilon}$ is not the conventional Kolmogorov 
length because $\overline{\epsilon}$ is not the Navier-Stokes energy dissipation rate 
$\epsilon = \nu \left<H_{1}\right>L^{-3}$. Instead we take an alternative 
route and use the estimate for $\left<H_{1}+\alpha^{2}H_{2}\right>$ from (\ref{uv5b}), which 
we repeat here
\bel{ad4}
\nu L^{-3}\left<H_{1} + \alpha^{2}H_{2}\right> 
\leq c\,\nu^{3}\ell^{-4}\Rey^{3}\,.
\ee
Thus 
\bel{ad5}
\overline{\epsilon} \leq c\,(L\lambda_{1}^{1/2})^{3}\nu^{3}\ell^{-4}\Rey^{3}\,,
\ee
which turns the result of Foias, Holm and Titi \cite{FHT2} into
\bel{ad6}
d_{F}(\mathcal{A}) \leq c\,\frac{V_{\alpha}V_{\ell}^{1/2}}{(L^{2}\lambda_{1})^{9/8}}\,\Rey^{9/4}\,,
\ee
where $L^{2}\lambda_{1} = 4\pi^{2}$. The right hand side blows up as $\alpha\to 0$ through 
$V_{\alpha}$. Despite this, the $\Rey^{9/4}$ estimate is, to our belief, the first time this 
has been achieved with this definition of $\Rey$. As has often been pointed out, this upper 
bound is also valid for the Hausdorff dimension $d_{H}(\mathcal{A})$ because 
$d_{H}(\mathcal{A})\leq d_{F}(\mathcal{A})$.

\section{\large A Theorem involving higher derivatives}\label{highderiv}

In terms of the number of degrees of freedom, the result in (\ref{ad6}) says that 
$\Rey^{3/4}\times\Rey^{3/4}\times\Rey^{3/4}$ resolution grid points are needed. 
However, as explained in \S\ref{intro}, what is not taken into account in attractor 
dimension estimates is the effect of strong dissipation-range intermittency where 
significant energy lies in wave-numbers larger than $\lambda_{k}^{-1}$. In this case 
estimates are needed for length-scales that are associated with higher derivatives. 
To obtain such estimates we begin by forming the combination 
\bel{Fdef1}
F_{n} = H_{n} +\tau^{2} \|\nabla^{n}\bdf\|_{2}^{2}\,,
\ee
where the quantity $\tau$ 
\bel{taudef}
\tau = \ell^{2}\nu^{-1}(\bG\ln\bG)^{-1/2}
\ee
where the $\ln\bG$-term is there for reasons explained in Appendix (\ref{App1.2}). 
We also define the combination
\bel{Jndef}
J_{n} = F_{n} + \alpha^{2}F_{n+1}\,.
\ee
The ultimate aim is to find time-averaged estimates for the $\kappa_{n,r}$ that appeared 
in (\ref{2nmom1}): this will be the subject of the next section. In preparation, we prove 
the following result: 
\begin{theorem}\label{thm1} As $\bG\to\infty$, for $n \geq 1$,~~$1\leq p \leq n$, $J_{n}$ 
satisfies
\bel{F1a}
\frac{dJ_{n}}{dt}\leq  -\shalf\nu\frac{J_{n}^{1+\frac{1}{p}}}{J_{n-p}^{1/p}} + 
c_{n,\alpha}\,\nu^{-1}\|\bu\|_{\infty}^{2}J_{n} + c_{1}\nu\ell^{-2}\Rey(\ln\Rey)J_{n}
\ee
and, for $n=0$,
\bel{F1a1}
\shalf\frac{dJ_{0}}{dt} \leq -\nu J_{1} + c_{1}\nu\ell^{-2}\Rey(\ln\Rey)J_{0}\,.
\ee
\end{theorem}
\par\medskip\noindent
\textbf{Proof:} The results on the pairing of $H_{0}$ and $H_{1}$ in (\ref{uv3}) and (\ref{uv4}) 
apply more generally:
\bel{uv6}
\I (\nabla^{n}\bu)\cdot(\nabla^{n}\bv)\,dV = 
\I \left\{|\nabla^{n}\bu|^{2} + \alpha^{2}|\nabla^{n+1}\bu|^{2}\right\}dV 
= H_{n} + \alpha^{2}H_{n+1}\,.
\ee
Thus
\bel{uv7}
\shalf\frac{d~}{dt}\left(H_{n} + \alpha^{2}H_{n+1}\right)
= \I(\nabla^{n}\bu)\cdot(\nabla^{n}\bv_{t})\,dV
\ee
which, from (\ref{uv2b}), gives the estimate
\beq{lad1}
\shalf\frac{d~}{dt}\left(H_{n} + \alpha^{2}H_{n+1}\right) &=&
\I(\nabla^{n}\bu)\cdot\left\{\nu\Delta\nabla^{n}\bv - 
\nabla^{n}(\bu\times\mbox{curl}\bv)\right\}\,dV\\
&\leq & -\nu \left(H_{n+1}+\alpha^{2}H_{n+2}\right) 
+ \left|\I (\nabla^{n+1}\bu)\left(\nabla^{n-1}(\bu\times\mbox{curl}\bv)\right)dV\right|\,.
\eeq
Upon separating the two constituent parts of $\bv = \bu -\alpha^{2}\Delta\bu$ within the last 
term in (\ref{lad1}), the first is found to satisfy
\beq{lad2}
\left|\I (\nabla^{n+1}\bu)\cdot\left(\nabla^{n-1}(\bu\times\mbox{curl}\bu)\right)\,dV\right| &\leq &
\|\bu\|_{\infty}H_{n+1}^{1/2}H_{n}^{1/2}\nonumber\\
&+ & H_{n+1}^{1/2}\sum_{m=1}^{n-1}C_{m}^{n-1}\|\nabla^{m}\bu\|_{p}\|\nabla^{n-m}\bu\|_{q}\,,
\eeq
where $p,~q$ must satisfy $p^{-1} + q^{-1} = 1/2$ according to H\"{o}lder's inequality. The 
first term on the RHS of (\ref{lad2}) is the $m=0$ term.  Now we use the two Gagliardo-Nirenberg inequalities
\bel{lad3}
\|\nabla^{m}\bu\|_{p}\leq c\,\|\nabla^{n}\bu\|_{2}^{a_{m}}\|\bu\|_{\infty}^{1-a_{m}}
\hspace{2cm}
\|\nabla^{n-m}\bu\|_{q}\leq c\,\|\nabla^{n}\bu\|_{2}^{b_{m}}\|\bu\|_{\infty}^{1-b_{m}}
\ee
where
\bel{lad4}
\frac{1}{p}-\frac{m}{d} = a_{m}\left(\frac{1}{2} - \frac{n}{d}\right)
\hspace{2cm}
\frac{1}{q}-\frac{n-m}{d} = b_{m}\left(\frac{1}{2} - \frac{n}{d}\right)\,.
\ee
Adding the two and noting that $p^{-1}+q^{-1}=\shalf$ implies that $a_{m} + b_{m} = 1$. In fact this
is true in $d$-dimensions. Thus the last term in (\ref{lad2}) is estimated by
\bel{lad5}
H_{n+1}^{1/2}\sum_{m=1}^{n-1}C_{m}^{n-1}\|\nabla^{m}\bu\|_{p}\|\nabla^{n-m}\bu\|_{q}
\leq c_{n}^{(1)}\,H_{n}^{1/2}H_{n+1}^{1/2}\|\bu\|_{\infty}\,.
\ee
We may approach the second constituent part of $\bv$ in the same manner 
as (\ref{lad2}). After an integration by parts we find
\bel{lad6}
\left|\I (\nabla^{n+2}\bu)\cdot(\nabla^{n-2}
\left(\bu\times\mbox{curl}(-\alpha^{2}\Delta\bu)\right)\,dV\right| 
\leq c_{n}^{(2)}\,\alpha^{2}H_{n+1}^{1/2}H_{n+2}^{1/2}\|\bu\|_{\infty}\,.
\ee
The estimate (\ref{lad1}) now beomes 
\beq{lad7a}
\shalf\frac{d~}{dt}\left(H_{n} + \alpha^{2}H_{n+1}\right) &\leq&
-\shalf \nu \left(H_{n+1} + \alpha^{2}H_{n+2}\right)
+ c_{n}\,\nu^{-1}\|\bu\|_{\infty}^{2}\left(H_{n} + \alpha^{2}H_{n+1}\right)\nonumber\\
&+& \|\nabla^{n}\bdf\|_{2}H_{n}^{1/2} + \alpha^{2}\|\nabla^{n+1}\bdf\|_{2}H_{n+1}^{1/2}\,.
\eeq
To turn this into an inequality for the $F_{n}$ we add and subtract to the negative terms 
on the right hand side, and break up the last terms to form the group of terms designated 
as $X_{n}$:
\bel{lad8}
X_{n} = \|\nabla^{n}\bdf\|_{2}H_{n}^{1/2} + \alpha^{2}\|\nabla^{n+1}\bdf\|_{2}H_{n+1}^{1/2}
+ \shalf\nu\ell^{-2}\tau^{2}\|\nabla^{n}\bdf\|_{2}^{2} + 
\shalf\nu\ell^{-2}\tau^{2}\alpha^{2}\|\nabla^{n+1}\bdf\|_{2}^{2}
\ee
with
\bel{lad7b}
\shalf\frac{d~}{dt}\left(F_{n} + \alpha^{2}F_{n+1}\right) \leq
-\shalf \nu \left(F_{n+1} + \alpha^{2}F_{n+2}\right)
+ c_{n}\,\nu^{-2}\|\bu\|_{\infty}^{2}\left(F_{n} + \alpha^{2}F_{n+1}\right) + X_{n}
\ee
where
\bel{lad9}
X_{n} \leq \left\{\frac{g}{2}H_{n} + \left(\frac{1}{2g\tau^2}+ 
\frac{\nu}{2\ell^2}\right)\tau^{2}\|\nabla^{n}\bdf\|_{2}\right\}
+ 
\left\{\frac{g}{2}H_{n+1} + \left(\frac{1}{2g\tau^2}+ 
\frac{\nu}{2\ell^2}\right)\tau^{2}\|\nabla^{n+1}\bdf\|_{2}\right\}\,.
\ee 
To make the coefficients of $H_{n}$ and $\tau^{2}\|\nabla^{n}\bdf\|_{2}$ equal,  
choose $g$ to satisfy
\bel{lad10}
g^{2} - \frac{g\nu}{\ell^2} - \frac{1}{\tau^2} = 0\,.
\ee
That is
\bel{lad11}
2g = \frac{\nu}{\ell^{2}} + \left[\left(\frac{\nu}{\ell^{2}}\right)^{2}+ \frac{4}{\tau^2}\right]^{1/2}\,.
\ee
Given $\tau$ in (\ref{taudef}) we have $g\approx \tau^{-1}$ as $\bG\to\infty$. Consequently, 
\bel{lad12}
\tau^{-1} = \ell^{2}\nu^{-1}(\bG\ln\bG)^{\shalf} \leq c\,\ell^{2}\nu^{-1}\Rey(\ln\Rey)
\ee
and
\bel{lad13}
X_{n} \leq \shalf \tau^{-1}\left(F_{n} + \alpha^{2}F_{n+1}\right)\,.
\ee
When applied to (\ref{lad7b}), the previous result yields
\bel{F1b}
\frac{dJ_{n}}{dt}\leq 
-\nu J_{n+1} + c_{n}\left(\nu^{-1}\|\bu\|_{\infty}^{2}+ \nu\ell^{-2}\Rey(\ln\Rey)\right)J_{n}
\ee
which is (\ref{F1a}). 
\par\medskip\noindent
The following Lemma deals with the $-J_{n+1}$-term in (\ref{F1b}):
\begin{lemma}\label{Jlemma} For $1\leq p \leq n$, the $J_{n}$ satisfy
\vspace{-2mm}
\bel{l1}
J_{n+1} \geq \frac{1}{2}\frac{J_{n}^{1+\frac{1}{p}}}{J_{n-p}^{1/p}}\,.
\ee
\end{lemma}
\textbf{Proof:} Firstly from (\ref{Fdef1}), by writing $F_{n}$ in Fourier transforms one finds
\bel{ft1}
F_{n} = H_{n} +\tau^{2} \|\nabla^{n}\bdf\|_{2}^{2} = \int k^{2n}|\hat{u}|^{2}\,dV_{k}\,,
\ee
where $|\hat{u}|^{2}$ includes the three components of the fluid velocity and the three components 
of the forcing. Using H\"{o}lder's inequality produces
\bel{ft1a}
F_{n} = \int \left(k^{2(n+q)}|\hat{u}|^{2}\right)^{\frac{p}{p+q}}
\left(k^{2(n-p)}|\hat{u}|^{2}\right)^{\frac{q}{p+q}}dV_{k} 
\leq F_{n+q}^{\frac{p}{p+q}}\,F_{n-p}^{\frac{p}{p+q}}\,.
\ee
Another application of H\"{o}lder's inequality also gives the standard result
\bel{l2}
N^{-p}\left(\sum_{i=1}^{N}a_{i}\right)^{p+1} \leq \sum_{i=1}^{N}|a_{i}|^{p+1}\,.
\ee
Now we find upper and lower bounds for the combination $F_{n}^{p+1} +\alpha^{2p+2}F_{n+1}^{p+1}$. 
Inequality (\ref{ft1a}) is used to find an upper bound
\beq{l3}
F_{n}^{p+1} + \alpha^{2p+2}F_{n+1}^{p+1} &\leq &
F_{n+1}^{p}F_{n-p} +  \alpha^{2p+2}F_{n+2}^{p}F_{n+1-p}\nonumber\\
&\leq& \left(F_{n+1} +\alpha^{2}F_{n+2}\right)^{p}
\left(F_{n-p} +\alpha^{2}F_{n+1-p}\right)\,.
\eeq
A lower bound comes from inequality (\ref{l2}) with $N=2$
\bel{l4}
F_{n}^{p+1} + [\alpha^{2}F_{n+1}]^{p+1} 
\geq 2^{-p}\left(F_{n} +\alpha^{2}F_{n+1}\right)^{p+1}\,.
\ee
Thus we have
\bel{l5}
2^{-p}J_{n}^{p+1} \leq J_{n+1}^{p}J_{n-p}
\ee
which gives the result.\hspace{9cm}$\square$
\par\medskip\noindent
The $n=0$ result (\ref{F1a1}) follows from (\ref{uv5a}) 
by using the same methods.\hspace{1.1cm}$\blacksquare$

\section{\large Length scales}

Now define the quantities $r <n$ (we take $r= n-p$)
\bel{ls1}
\kappa_{n,r}(t) = \left(\frac{J_{n}}{J_{r}}\right)^{\frac{1}{2(n-r)}}\,,
\ee
which act as $t$-dependent wave-numbers and thus have the dimension of inverse length scales. 
For $r=0$ they are analogous to the $2n$-th moments of the energy spectrum. In what follows 
we find upper bounds for $\left<\kappa_{n,0}^2\right>$. These are bounds on an infinite 
series of inverse squared length scales.
\par\medskip\noindent
In Theorem 1 for $r<n$, dividing by the $J_{n}$ \& time averaging , we have 
\beq{ls2}
\left<\kappa_{1,0}^2\right> &\leq& c_{1}\ell^{-2}\Rey\ln\Rey\,,\\
\left<\kappa_{n,r}^2\right> &\leq& c_{n,\alpha}\,\nu^{-2}\left<\|\bu\|_{\infty}^{2}\right> 
+ c_{1}\ell^{-2}\Rey\ln\Rey\,.
\eeq
The results for this can be summarized in the following:
\begin{theorem}\label{thm2}
As $\bG\to\infty$, the four time averages are estimated as
\beq{th2a}
\ell^{2}\left<\kappa_{n,0}^2\right> &\leq& c_{n,\alpha}V_{\alpha}^{\frac{n-1}{n}}
\Rey^{\frac{11}{4} - \frac{7}{4n}}(\ln\Rey)^{\frac{1}{n}} + c_{1}\Rey\ln\Rey\,,\label{th2a}\\
\ell^{2}\left<\kappa_{1,0}^2\right> &\leq& c_{1}\Rey\ln\Rey\,,\label{th2b}\\
\ell^{2}\nu^{-2}\left<\|\bu\|_{\infty}^2\right> &\leq& c\,V_{\alpha}\,\Rey^{11/4}\,,\label{th2c}\\
\alpha\ell\nu^{-1}
\left<\|\nabla\bu\|_{\infty}\right> &\leq& c\,V_{\alpha}^{1/4}V_{\ell}^{1/2}\,\Rey^{35/16}\,.
\label{th2d}
\eeq
\end{theorem}
\par\medskip\noindent
\textbf{Remark:} These four results are also listed in the Table of \S\ref{intro}.
\par\medskip\noindent
\textbf{Proof:} Agmon's inequality says that
\bel{ls3a}
\|\bu\|_{\infty}^{2}\leq c\,\|\nabla^{2}\bu\|_{2}\|\nabla\bu\|_{2} = 
c\,H_{2}^{1/2}H_{1}^{1/2}\,.
\ee
Moreover, we also notice that because $H_{1}\leq H_{2}^{1/2}H_{0}^{1/2}$ and 
$U = L^{-3/2}\left<H_{0}\right>^{1/2}$ we have 
\bel{ls3b}
\left<\|\bu\|_{\infty}^{2}\right>^{2} \leq 
c\,\left<H_{0}\right>^{1/2}\left<H_{2}\right>^{3/2}
\leq c\,L^{3/2}U\left<H_{2}\right>^{3/2}\,.
\ee
which simplifies to\footnote{Herein lies the difference between $3D$ \lans equations and 
the $3D$ Navier-Stokes equations: in the former we have a bound 
on $\left<H_{2}\right>$ from (\ref{uv5b}) which is missing in the latter.}
\bel{ls3c}
\ell^{2}\nu^{-2}\left<\|\bu\|_{\infty}^2\right> \leq c\,V_{\alpha}\,\Rey^{11/4}
\ee
which is (\ref{th2c}). With the dimensionless volumes $V_{\alpha}$ and $V_{\ell}$ defined 
in (\ref{vol}), we have
\beq{ls5a}
\ell^{2}\left<\kappa_{1,0}^2\right> &\leq& c_{1}\Rey\ln\Rey\,,\\
\ell^{2}\left<\kappa_{n,r}^2\right> &\leq& c_{n,\alpha}V_{\alpha}\Rey^{11/4}
+ c_{1}\Rey\ln\Rey\,.\label{ls5b}
\eeq
For $r=0$ we can improve this by writing
\beq{ls6}
\left<\kappa_{n,0}^2\right> &=&\left<\left(\frac{F_{n}}{F_{1}}\right)^{1/n}
\left(\frac{F_{1}}{F_{0}}\right)^{1/n}\right>\nonumber\\
&=& \left<(\kappa_{n,1}^{2})^{\frac{n-1}{n}}(\kappa_{1,0}^{2})^{\frac{1}{n}}\right>\nonumber\\
&\leq& \left<\kappa_{n,1}^{2}\right>^{\frac{n-1}{n}}\left<\kappa_{1,0}^{2}\right>^{\frac{1}{n}}\,,
\eeq
and then using the estimates in (\ref{ls5a}) and (\ref{ls5b}) for $n \geq 1$, which gives
\bel{imp1a}
\ell^{2}\left<\kappa_{n,0}^2\right> \leq c_{n,\alpha}V_{\alpha}^{\frac{n-1}{n}}
\Rey^{\frac{11}{4} - \frac{7}{4n}}(\ln\Rey)^{\frac{1}{n}} + c_{1}\Rey\ln\Rey\,.
\ee
This is (\ref{th2a}). Note that when $n=1$ we return to $\ell^{2}\left<\kappa_{n,0}^2\right> 
\leq c_{1}\Rey\ln\Rey$. 
\par\medskip\noindent
It is also possible to estimate $\left<\|\nabla\bu\|_{\infty}\right>$
\bel{gradu1}
\left<\|\nabla\bu\|_{\infty}\right> \leq c\,\left<H_{3}^{1/4}H_{2}^{1/4}\right>
= c\,\left<\kappa_{3,2}^{1/2}H_{2}^{1/2}\right>
\leq c\,\left<\kappa_{3,2}^{2}\right>^{1/4}\left<H_{2}\right>^{1/2}\,.
\ee
Using (\ref{uv5b}) and (\ref{ls5b}) we find
\bel{gradu2}
\alpha\ell\nu^{-1}
\left<\|\nabla\bu\|_{\infty}\right> \leq c\,\left<H_{3}^{1/4}H_{2}^{1/4}\right>
\leq c\,V_{\alpha}^{1/4}V_{\ell}^{1/2}\,\Rey^{35/16}\,,
\ee
which is (\ref{th2d}). \hspace{11cm}$\blacksquare$
\par\medskip\noindent
\textbf{Acknowledgements:} We are enormously grateful to C. Foias and E. S. Titi for many illuminating discussions of the \lans equations and their analytical properties. We also thank C. R. Doering for encouraging explanations of the results of \cite{DF}. The work of DDH was partially supported by US DOE, Office of Science, Applied Mathematics program of the Mathematical, Information, and Computational Sciences Division (MICS).

\appendix
\section{\large Issues concerning the forcing}

\subsection{Bounds concerning $\bG$ and $\Rey$}\label{App1.1}

Doering and Foias \cite{DF} split the forcing function $\bdf(\bx)$ into its magnitude $F$ 
and its ``shape'' $\bphi$ such that
\bel{p1}
\bdf(\bx) = F\bphi(\ell^{-1}\bx)
\ee
where $\ell$ is the longest length scale in the force. On the unit torus $\mathbb{I}_{d}$, 
$\bphi$ is a mean-zero, divergence-free vector field with the chosen normalization 
property 
\bel{p1a}
\int_{\mathbb{I}_{d}} \left|\nabla^{-1}_{y}\bphi\right|^{2}\,d^{d}y = 1\,.
\ee
$L^{2}$-norms of $\bdf$ on $\mathbb{I}^{d}$ are 
\bel{p2}
\|\nabla^{N}\bdf\|_{2}^{2} = C_{N}\ell^{-2N} L^{d} F^{2}
\ee
where the coefficients $C_{N}$ refer to the shape of the force but not its magnitude
\bel{p3}
C_{M} = \sum_{n} \left| 2\pi n\right|^{2N} |\hat{\bphi}_{n}|^{2}\,.
\ee
Doering and Foias \cite{DF} showed that various bounds exist such as (among others)
\bel{p4}
\|\nabla\Delta^{-M}\bdf\|_{\infty} = D_{M}F \ell^{2M-1}\,.
\ee
The energy dissipation rate $\epsilon$ is  
\bel{p5}
\epsilon = \left<\nu L^{-d}\I |\nabla\bu|^{2}\,dV\right> = \nu L^{-d}\left<H_{1}\right>\,.
\ee
In terms of $F$ the Grashof number in (\ref{f1}) becomes
\bel{Grdef}
\bG = F\ell^{3}/\nu^{2}
\ee
and the Taylor micro-scale $\lambda_{T}$ is related to $U$ via $\lambda_{T} = \sqrt{\nu U^{2}/\epsilon}$\,,
which is consistent with the definition $\lambda_{T}^{-2} = \left<H_{1}\right>/\left<H_{0}\right>$.
\par\bigskip\noindent
The \lans equations (\ref{uv2a}) and (\ref{uv2b}) can also formally be re-written as
\bel{uv2c}
\bu_{t}+\bu\cdot\nabla\bu -\nu\Delta\bu + \nabla\tilde{p} = (1-\alpha^{2}\Delta)^{-1}
\left\{\bdf(\bx) + \alpha^{2}\mbox{div}\mathbb{T}\right\} 
\ee
where the tensor $\mathbb{T}$ is defined as
\bel{Tdef}
\mathbb{T} = \nabla\bu\cdot\nabla\bu + \nabla\bu\cdot\nabla\bu^{T} 
- \nabla\bu^{T}\cdot\nabla\bu\,.
\ee
Following the procedure in \cite{DF} (pg 296 equation (2.9)) and multiplying 
by $(-\Delta^{-M})\bdf$ we have 
\beq{a1}
\frac{d~}{dt}\int_{\mathbb{I}_{d}} \bu\cdot[(-\Delta^{-M})\bdf]\,dV 
&=& - \nu\int_{\mathbb{I}_{d}} \Delta\bu\cdot[(-\Delta^{-M})\bdf] 
- \int_{\mathbb{I}_{d}} \bu\cdot\nabla\bu \cdot[(-\Delta^{-M})\bdf]\,dV \nonumber\\
&+&\int_{\mathbb{I}_{d}} [(-\Delta^{-M})\bdf]\cdot (1-\alpha^{2}\Delta)^{-1}
\left\{\bdf + \alpha^{2}\mbox{div}\mathbb{T}\right\}\,dV\,.
\eeq
Now there are two strategies:
\par\medskip\noindent
\textbf{1) To prove that $\bG \leq c\,\Rey^{2}$:} integrate all the terms by parts, 
and take the time average
\beq{a2}
\left<L^{-d}\int_{\mathbb{I}_{d}} \nabla^{-M}\bdf\cdot (1-\alpha^{2}\Delta)^{-1}
\nabla^{-M}\bdf \,dV\right> &\leq& 
\left<L^{-d}\nu\int_{\mathbb{I}_{d}}\bu\cdot[(-\Delta^{-M+1})\bdf]\,dV\right>\nonumber\\
&-&\left<L^{-d}\int_{\mathbb{I}_{d}}\bu\cdot[\nabla[(-\Delta^{-M})]\bdf]\cdot\bu\,dV\right>\\
&+&\left<L^{-d}\int_{\mathbb{I}_{d}}\left|\nabla[(-\Delta^{-M})\bdf]\right|
\left|\frac{\alpha^{2}}{(1-\alpha^{2}\Delta)}
\mathbb{T}\right|\,dV\right>\nonumber
\eeq
Using the scaling properties of $\bphi$ and a Fourier transform on the last term
\beq{a3}
L^{-d}\int_{\mathbb{I}_{d}}\left|\nabla[(-\Delta^{-M})\bdf]\right|\left|\frac{\alpha^{2}}{(1-\alpha^{2}\Delta)}
\mathbb{T}\right|\,dV
&\leq& D_{M}F\ell^{2M-1}\int \frac{\alpha^{2}k^{2}}{1+ \alpha^{2}k^{2}}|\hat{\bu}|^{2}\,dV_{k}\nonumber\\
&\leq& D_{M}F\ell^{2M-1}\int|\hat{\bu}|^{2}\,dV_{k}
\eeq
where (\ref{p4}) defines $D_{M}$. Thus (\ref{a2}) turns into 
\bel{a4}
c_{0}\frac{F^{2}\ell^{2M}}{1+\alpha^{2}\ell^{-2}}
\leq c_{1}\nu F\ell^{2M-2}U + c_{2}\ell^{2M-1}F U^{2}\,,
\ee
where the $U^2$-term contains the contributions from both nonlinear terms and the constants (not 
explicitly given) contain the shape of the body forcing. Using (\ref{Grdef}), (\ref{a4}) becomes 
\bel{a5}
\bG \leq c\,\left(\Rey + \Rey^{2}\right)\,,\hspace{3cm}\bG \to\infty\,,
\ee
the only difference from the Navier-Stokes equations being the value of the constant.
\par\medskip\noindent
\textbf{2) To prove that $\epsilon \geq c\,\nu^{3}\ell^{-3}L^{-1}\bG$:} return to (\ref{a1}) and 
take a different route. Firstly in the Laplacian term use one derivative on $\bu$ and another 
on the forcing. Then keep the $\bu\cdot\nabla\bu$ advection term. Finally integrate the 
$\mathbb{T}$-term by parts 
and exploit the fact that $(1-\alpha^{2}\Delta)^{-1}$ is a symmetric operator. 
\beq{a6}
\left<L^{-d}\int_{\mathbb{I}_{d}} (\nabla^{-M}\bdf)
\cdot\frac{1}{(1-\alpha^{2}\Delta)}\nabla^{-M}\bdf \,dV\right>
&\leq& 
\left<L^{-d}\nu\int_{\mathbb{I}_{d}} |\nabla\bu|\cdot[(\nabla\Delta^{-M})\bdf]\right> \nonumber\\
&+& \left<L^{-d}\int_{\mathbb{I}_{d}} |\bu|\,|\nabla\bu|\,|(-\Delta^{-M})\bdf|\,dV\right>\\
&+& \left<L^{-d}\int_{\mathbb{I}_{d}}\left|\frac{\alpha^{2}}{(1-\alpha^{2}\Delta)}
\nabla[(-\Delta^{-M})\bdf]\right|\left|\mathbb{T}\right|\,dV\right>\,.\nonumber
\eeq
Thus (\ref{a6}) turns into 
\bel{a7}
c_{0}\frac{F^{2}\ell^{2M}}{1+\alpha^{2}\ell^{-2}}\leq c_{1}F\nu^{1/2}\ell^{2M-1}\epsilon^{1/2}+
c_{2}F\nu^{-1/2}\ell^{2M}\epsilon^{1/2}U + c_{3}\alpha^{2}\ell^{2M-1}F\nu^{-1}\epsilon \,.
\ee 
We assume the ordering of length scales as $L \geq \ell \geq \alpha \geq \lambda_{T}$ and use 
the fact that $U^{2} = \lambda_{T}^{2}\epsilon\nu^{-1} \leq L^{2}\epsilon\nu^{-1}$. Then 
(\ref{a7}) becomes 
\bel{a8}
\shalf c_{0} F \leq c_{1}\nu^{1/2}\ell^{-1}\epsilon^{1/2} + \nu^{-1}\epsilon\,
(c_{2}L + c_{3} \alpha^{2}\ell^{-1}) \leq c_{1}\nu^{1/2}\ell^{-1}\epsilon^{1/2} + 
c_{4}\nu^{-1}L\epsilon \,.
\ee
For $\bG \to \infty$ the last term on the RHS is the dominant one: we have
\bel{a9}
\epsilon \geq c_{5}\,\ell^{-3}L^{-1} \nu^{3}\bG
\ee
which, with a different constant, agrees with the result in Doering and Foias \cite{DF}. This 
inequality is used in the next subsection.

\subsection{Forcing \& the fluid response}\label{App1.2}

For technical reasons, we must address the possibility that in their evolution the quantities 
$H_{n}$ might take small values. Thus we need to circumvent problems that may arise when dividing 
by these (squared) semi-norms. We follow Doering and Gibbon \cite{DG02} who introduced the modified
quantities in (\ref{Fdef1})
\bel{Fndef}
F_{n} = H_{n} + \tau^{2}\|\nabla^{n}\bdf\|^{2}_{2}
\ee
where the ``time-scale'' $\tau$ is to be chosen for our convenience. So long as
$\tau \neq 0$, the $F_{n}$ are bounded away from zero by the explicit value
$\tau^{2}L^{3}\ell^{-2n}f_{rms}^2$. Moreover, we may choose $\tau$ to depend on 
the parameters of the problem such that $\left<F_{n}\right>\sim
\left<H_{n}\right>$ as $\bG\to\infty$. To see how to achieve this, let us define
\bel{add1}
\tau = \ell^{2}\nu^{-1}(\bG\ln\bG)^{-1/2}\,.
\ee
Then the additional term in (\ref{Fndef}) is
\begin{eqnarray}\label{add2}
\tau^{2}\|\nabla^{n}\bdf\|_{2}^{2} &=& 
L^{3}\nu^{-2}\ell^{4-2n}f_{rms}^{2}(\bG\ln\bG)^{-1}\nonumber\\
&=& \nu^{2}\ell^{-(2n+2)}L^{3}\bG(\ln\bG)^{-1}
\end{eqnarray}
Recalling the {\em a priori} bound on the far right hand side of (\ref{a9})
\begin{eqnarray}\label{add3}
\tau^{2}\|\nabla^{n}\bdf\|_{2}^{2} 
&\leq & c_{6}\epsilon\,\ell^{-(2n-1)}L^{4}\nu^{-1}(\ln\bG)^{-1}\nonumber\\
& = & c_{6}\left(\frac{L}{\ell}\right)^{(2n-1)}
L^{-2(n-1)}\bigl<H_{1}\bigr>(\ln\bG)^{-1}
\end{eqnarray}
Using Poincar\'{e}'s inequality in the form 
$H_{1} \leq (2\pi L)^{2(n-1)}H_{n}$, as $\bG\to\infty$ we have
\bel{}
\frac{\tau^{2}\|\nabla^{n}\bdf\|_{2}^{2}}{\bigl<H_{n}\bigr>} \leq
c_{6}\left(\frac{L}{\ell}\right)^{(2n-1)}(\ln\bG)^{-1}
\ee
Hence, the additional forcing term in (\ref{Fndef}) 
becomes negligible with respect to $\left<H_{n}\right>$ as $\bG\to \infty$, so the 
forcing does not dominate the response.


\end{document}